# Evidence for Weyl fermions in the elemental semiconductor tellurium

Nan Zhang[1,2,#], Gan Zhao[1,#], Lin Li[1,2,#,*], Pengdong Wang[3], Lin Xie[4], Hui Li[5], Zhiyong Lin[1,2], Jiaqing He[4], Zhe Sun[3], Zhengfei Wang[1,*], Zhenyu Zhang[1], and Changgan Zeng[1,2,*]

[1] *International Center for Quantum Design of Functional Materials, Hefei National Laboratory for Physical Sciences at the Microscale, and Synergetic Innovation Center of Quantum Information & Quantum Physics, University of Science and Technology of China, Hefei, Anhui 230026, China*

[2] *CAS Key Laboratory of Strongly Coupled Quantum Matter Physics, Department of Physics, University of Science and Technology of China, Hefei, Anhui 230026, China*

[3] *National Synchrotron Radiation Laboratory, University of Science and Technology of China, Hefei, Anhui 230029, China*

[4] *Department of Physics, Southern University of Science and Technology, Shenzhen 518055, China*

[5] *Institutes of Physical Science and Information Technology, Anhui University, Hefei, Anhui 230601, China*

[#] These authors contributed equally to this work.

[*] Correspondence and requests for materials should be addressed to C. Z. (email: cgzeng@ustc.edu.cn), Z. W. (email: zfwang15@ustc.edu.cn), and L. L. (email: lilin@ustc.edu.cn).

**The recent discovery of Weyl fermions in solids enables exploitation of relativistic physics and development of a spectrum of intriguing physical phenomena[1-9]. They are constituted of pairs of Weyl points with two-fold band degeneracy[1,2,10], which in principle can be hosted in any materials without inversion or time-reversal symmetry. However, previous studies of Weyl fermions have been limited exclusively to semimetals. Here, by combining magneto-transport measurements, angle-resolved photoemission spectroscopy, and band structure calculations, Weyl fermions are identified in an elemental semiconductor tellurium. This is mainly achieved by direct observation of the representative transport signatures of the chiral anomaly, including the negative longitudinal magnetoresistance[6,9,11-17] and the planar Hall effect[15,18-23]. Semiconductor materials are well suited for band engineering, and therefore provide an ideal platform for manipulating the fundamental Weyl fermionic behaviors. Furthermore, introduction of Weyl physics into semiconductors to develop "Weyl semiconductors" also creates a new degree of freedom for the future design of semiconductor electronic and optoelectronic devices.**

The nontrivial Weyl band topology has noteworthy high-energy physics implications, resulting in many exotic transport phenomena, such as chiral-anomaly-induced negative longitudinal magnetoresistance[6,9,11-17] and planar Hall effect[15,18-23], giant anomalous Hall effect[8,9], and unconventional quantum oscillations effect[24]. Generally, transport signals are determined by the band structure around the Fermi level. Hence, novel Weyl-related physics was considered to be the unique feature of semimetal materials, where the Weyl points are located near the Fermi level. However, recent theoretical work demonstrates that chiral-anomaly-induced transport phenomena can persist even when the Weyl points are located far from the Fermi level[25]. This suggests that Weyl-related exotic phenomena need not be restricted to semimetals. Physically, Weyl points represent band crossings with two-fold degeneracy, and they could in principle exist in any material lacking inversion or time-reversal symmetry.

Bearing this in mind, a natural next step in Weyl-physics-related progress is to extend Weyl fermions into semiconductors, a move which has been overlooked in previous research. Semiconductors are particularly suitable for Weyl devices due to their high tunability and compatibility with modern electronic industry. In a narrow gap semiconductor with no inversion symmetry, the spin-orbit coupling (SOC) enables band splittings and crossings in its bulk conduction bands (CBs) or valence bands (VBs)[26], which may result in the formation of Weyl points. If these Weyl points are located near the top of the VBs (or the bottom of the CBs), chiral-anomaly-induced exotic phenomena are expected. Here, semiconductors hosting Weyl fermions can be referred to as "Weyl semiconductors", which is schematically shown in Fig. 1a. In this work, which leverages experimental measurements and theoretical calculations, tellurium (Te) is identified as a realistic material system of "Weyl semiconductor". As an elemental semiconductor with relatively high chemical stability, it provides a simple and ideal platform for the exploration of Weyl physics in material systems beyond semimetals.

The helical structure of Te with no inversion symmetry is shown in Fig. 1b. The first principles band structure of Te is shown in Supplementary Fig. S1 (see Methods for calculation details), which demonstrates that Te is a narrow gap semiconductor exhibiting strong SOC. Its band gap (~0.38 eV) is near the corner of the Brillouin zone, i.e., the H point (Fig. 1b), which is consistent with previous results[27,28]. Along the high symmetric H-L line, two Weyl points arise from the crossing of two spin-splitting hole bands, as shown in Fig. 1c. They are located below the Fermi level by -0.20 eV (designated $W_1$) and -0.36 eV (designated $W_2$), respectively. Therefore, the theoretical results support the existence of Weyl fermions in the top VBs of Te.

In order to confirm this theoretical prediction, angle-resolved photoemission spectroscopy (ARPES) measurements were taken on the cleaved $(11\bar{2}0)$ surface of the Te crystals (see Methods for the measurement details). As shown in Fig. 1d, there is ARPES intensity at H point on the Fermi surface map for the $k_y$-$k_z$ plane with $k_x \sim \pi$, which comes from the top of VB of Te crystals. Furthermore, when the theoretically calculated bands are overlaid with the ARPES intensity along the H-L line (cut I in Fig. 1d), there is good agreement between the data sets in a large energy window (Fig. 1e). Due to limited ARPES resolution, it is challenging to clearly identify these two Weyl points ($W_1$ and $W_2$) in our measurements. However, the striking consistency between the theoretical and experimental results strongly indicates the existence of Weyl points in the top VBs of Te, which inspired us to further study the intriguing Weyl-related transport phenomena in Te.

A material system hosting Weyl fermions is expected to exhibit the so-called chiral anomaly[5]. The key transport signature of such a phenomenon is the emergence of negative longitudinal magnetoresistance (NLMR) in the presence of parallel magnetic and electric fields[6,9,11-17]. To exploit this, magneto-transport measurements were systematically performed. Below we will present the comprehensive results for sample #3. Similar behaviors are also observed in the other crystal samples (See Supplementary Figs. S3 and S4).

Given the rod-like shape of Te crystals (inset of Fig. 2a), the current is applied along the long axis (the [0001] direction) during all transport measurements conducted in this study. Fig. 2a shows the temperature ($T$)-dependent resistance ($R$) curve, from which the typical transport behaviors of the doped semiconductor are observed (as detailed in Supplementary Note 2 and Supplementary Fig. S5a). Figs. 2b and 2c show the curves for the magnetoresistance (MR, defined as $(R(B) - R(0))/R(0) \times 100\%$) across a temperature range of 25-100 K under perpendicular and parallel fields, respectively. Further insights into the temperature dependence of the MR properties (in a wide temperature range of 2-100 K) are presented in Supplementary Fig. S5 and discussed in Supplementary Note 3. As shown in Fig. 2b, the MR curves exhibit clear linear behaviors at various temperatures when the magnetic field ($B$) is perpendicular to the current ($I$), similar to previous observations in doped semiconductors possessing small Fermi surface pockets such as InSb[29]. Such character may be reasonably explained with the quantum linear model proposed by Abrikosov[30], when all carriers are condensed into the lowest Landau levels. For $B \parallel I$, the MR is positive for the small field regime, and then switches to a negative value at a critical field $B_C$ (Fig. 2c). The magnitude of negative MR increases monotonically with increasing $B$ up to 14 T. For T=25 K, the magnitude reaches to ~22% at 14 T, which is comparable to the NLMR effects reported for several topological semimetals such as $WTe_2$[14] and $Co_3Sn_2S_2$[9]. As the temperature increases, $B_C$ shifts to a higher value and the negative MR effect degrades accordingly due to the thermal effect, but remains observable up to ~100 K. The occurrence of the NLMR in Te may be related to the chiral anomaly, as discussed below.

Theoretically, the longitudinal magnetoconductivity, including the contribution from the chiral anomaly, can be described using the following formula[11]:

$$\sigma(B) = (1 + C_w B^2) \cdot \sigma_{WAL} + \sigma_N \tag{1}$$

where $C_w$ is a positive parameter representing the chiral anomaly contribution to the conductivity, and $\sigma_{WAL}$ and $\sigma_N$ are the conductivities imparted by the weak anti-localization effect and conventional non-linear band contributions around the Fermi level, respectively. Here $\sigma_{WAL} = \sigma_0 + a\sqrt{B}$, $\sigma_N^{-1} = \rho_0 + A \cdot B^2$, and $\sigma_0$ is zero-field conductivity.

The inset in Fig. 2d shows a typical plot for the field dependence of the longitudinal conductivity at 25 K. The experimental data is fitted with formula (1), which is clearly in excellent agreement. By applying this analysis to the MR data for a range of temperatures (see Supplementary Fig. S7a), we extract a trend of monotonically increasing $C_w$ with decreasing temperature down to 15 K (Fig. 2d). It is noted that the chiral coefficient $C_w$ can be well fitted via the formula $C_w \propto v_F^3 \tau_v / (T^2 + \mu^2/\pi^2)$[13], the theoretically expected temperature-dependent property for the chiral anomaly, where $v_F$ is the Fermi velocity, $\tau_v$ is the chirality-changing scattering time, and $\mu$ is the chemical potential relative to the Weyl points.

The excellent agreement between the experimental data and the theoretical fittings strongly supports that the NLMR effect observed in Te originates from the chiral anomaly. To gain further insight into the observed negative MR, we also study its angular-dependent character. Fig. 2e shows the MR for different angles ($\theta$) of $B$ with respect to *I* as measured at 25 K. When the magnetic field rotates away from *B*⊥*I*, the positive MR magnitude decreases accordingly. Signatures of negative MR occur for $\theta < 25°$, peaking at $\theta \sim 0°$ (*B*∥*I*). In addition, the $C_w$ obtained from the fitted data reveals linear dependence of $\cos^2\theta$ (Fig. 2f, see also Supplementary Fig. S7b), which is consistent with observations in topological semimetals[15]. Similar angular-dependent MR behavior is observed at temperatures down to 2 K, although the negative MR is largely outweighed by the enormous positive MR background (Supplementary Fig. S8).

Thus far, the negative MR observed in Te crystals under *B*∥*I* could be well explained via the chiral anomaly, with the behavior in good agreement with previous studies of Weyl semimetals. Other possible origins, such as weak localization and current jetting, have also been examined carefully and eventually ruled out (see Supplementary Note 4).

Planar Hall effect (PHE), which manifests as the appearance of in-plane transverse voltage when the in-plane magnetic field is not exactly parallel or perpendicular to the current, is another important transport signature of the chiral anomaly[22,23]. Compared to the NLMR effect, the PHE is less sensitive to spin scattering and is therefore regarded as a more reliable experimental tool for probing the chiral anomaly in topological semimetals[15,18-21]. Theoretically, the PHE in topological semimetals can be described by the following formulas[22,23]:

$$\rho_{xx} = \rho_\perp - \Delta\rho^{chiral} \cos^2\varphi \qquad (2)$$

$$\rho_{xy} = -\Delta\rho^{chiral} \sin\varphi \cos\varphi \qquad (3)$$

where $\Delta\rho^{chiral} = \rho_\perp - \rho_\parallel$ is the anisotropic resistivity originated from the chiral anomaly, $\rho_\perp$ and $\rho_\parallel$ are the resistivities corresponding to the magnetic field perpendicular to and along the current direction, respectively, and $\varphi$ is the angle between the in-plane magnetic field and the current.

The configuration for measuring the PHE is depicted in the inset of Fig. 3a. A conventional Hall component may still exist due to misalignment between the actual rotation plane and the sample plane, but it can be eliminated by simply averaging the

Hall resistances measured under positive and negative magnetic fields. Fig. 3a shows the symmetrized planar Hall resistivity $\rho_{xy}$ versus $\varphi$ measured at 25 K over a range of magnetic fields. The in-plane anisotropic magnetoresistance (AMR, defined as $(R(\varphi) - R(\varphi = 90°))/R(\varphi = 90°) \times 100\%$) was measured simultaneously, and the results are plotted in Supplementary Fig. S9a. Both the planar Hall resistivity and the in-plane AMR display a 180° periodic angular dependence. AMR reaches its maximum and minimum at 90° and 0°, respectively, while $\rho_{xy}$ experiences its maximum and minimum at 135° and 45°, respectively. These observations align well with the typical characters of the PHE.

From a practical standpoint, it is important to consider that a small longitudinal resistivity component can contribute to the planar Hall resistivity due to misalignment during fabrication of the Hall bar device. Thus, formula (3) should be modified to achieve more precise fitting of the experimental $\rho_{xy}$. The revised equation is[20]

$$\rho_{xy} = -\Delta\rho^{chiral} \sin\varphi \cos\varphi + b\,\Delta\rho^{chiral} \cos^2\varphi + c \qquad (4)$$

where the first term is the intrinsic PHE due to the chiral anomaly, and the second and third terms are the in-plane AMR and longitudinal offset caused by the Hall configuration misalignment. Fig. 3a shows that equation (4) provides an excellent fit for the experimental $\rho_{xy}$ across a range of magnetic fields, further strengthening the case for the existence of chiral anomaly in Te crystals.

For a more quantitative PHE analysis, we extract the field dependent $\Delta\rho^{chiral}$ at 25 K. The plot in Fig. 3c clearly demonstrates that the obtained $\Delta\rho^{chiral}$ increases monotonically with increasing magnetic field. Different $B$ dependencies are revealed in different field regimes. For $B < 3$ T, $\Delta\rho^{chiral}$ is proportional to $B^2$. In the relatively high field regime ($B > 4.5$ T), $\Delta\rho^{chiral}$ exhibits nearly linear variation with $B$ and displays no evidence of saturation up to $B = 14$ T. Similar field dependence has been previously reported for the PHE in topological semimetals[20,21].

Fig. 3b shows the temperature-dependent PHE behavior (see Supplementary Fig. S9b for the AMR data taken simultaneously). With increasing temperature, the extracted

$\Delta \rho^{chiral}$ decreases accordingly and becomes negligible at $T \sim 100$ K (Fig. 3d). Such behavior is remarkably consistent with that of the $C_w$ shown in Fig. 2d, suggesting the same underlying physical origin of both the PHE and NLMR in the present Te crystals (see Supplementary Note 3 for more discussion).

NLMR and PHE are the two prominent transport signatures of the chiral anomaly. The simultaneous observation of these two effects in Te crystals definitively demonstrates that semiconductors can also host Weyl fermions. Recent theoretical studies predicted that Te can be turned into a Weyl semimetal via closing the band gap by applying high pressure[28,31]. In contrast, here we reveal that the Weyl fermions and related exotic transport properties could be directly realized in Te without gap closing. Note that the conductivity contributed by the chiral anomaly is predicted to be inversely proportional to the square of chemical potential relative to the Weyl point[11,16], the NLMR effect thus could be modulated via changing the carrier density. For Te, higher carrier density means a smaller separation between $E_F$ and the Weyl point, and the corresponding magnitude of the NLMR effect is expected to be enhanced accordingly. Via comparison of the chiral coefficient $C_w$ for samples of various carrier densities, we indeed observe such a tendency in Te crystals experimentally, as shown in Fig. 4a (see Supplementary Fig. S4 and Supplementary Note 5 for more details). Similar results have been reported for typical Weyl semimetals[6,12,15], thus providing further compelling evidence for the topological nature of Te crystals.

The present work clearly demonstrates the realization of Weyl fermions and Weyl-related properties in narrow-gap semiconductors with strong SOC and without inversion symmetry, thus greatly broadening the scope of topological materials. Furthermore, in such Weyl semiconductors, the highly tunable electronic performance enables further manipulation of Weyl fermions through various means commonly adopted in semiconductor electronics, such as electrostatic gating and optical illumination. For example, tuning the Fermi level from the VBs (or CBs) into the band gap induces a metal to insulator transition. At the same time, a corresponding transition from a topological nontrivial state to a trivial state occurs, as schematically illustrated in Fig. 4b. This simultaneous switch of both electrical conductivity and the chiral anomaly is unattainable in Weyl semimetals.

Moreover, the successful introduction of Weyl physics into semiconductor systems may also offer a new dimension for the future design of semiconductor devices. Recently, giant photovoltaic effect in Weyl semimetals has been demonstrated, originated from the optical selection rule of chiral Weyl points[32,33]. Such chiral control can also enrich the already fascinating optoelectronic properties of semiconductors, and therefore be utilized to design novel topological optoelectronic devices based on Weyl semiconductors. In addition, the existence of Weyl fermions in semiconductors ensuring robust chiral transport against external perturbations[7], also creates new opportunities for the development of low-consumption semiconductor electronic devices.

**Methods**

**Growth of Te single crystals.** High-quality Te single crystals were grown using a physical vapor transport technique. High-purity (99.999% pure) Te powder was loaded into a quartz tube, and a small amount of C powder was added to remove trace oxygen. The quartz tube was heated to 1000 ˚C over the course of 5 hours and subsequently maintained at 1000 ˚C for 1 hour. Then the tube was cooled at a rate of 20 ˚C/h to 400 ˚C and 300 ˚C for the hot and the cold zone, respectively. After maintaining the set temperature for 2 weeks, the tube was slowly cooled to room temperature, and rod-like silvery crystals (typical dimensions of $5 \times 0.3 \times 0.1$ mm$^3$) were obtained (see the inset of Fig. 2a).

**Structure and composition characterizations.** The structure of Te single crystals was measured by a Rigaku SmartLab X-ray diffractometer (XRD) at room temperature using monochromic Cu-Kα radiation (λ=1.5418 Å). A typical XRD spectrum is shown in Supplementary Fig. S2a. In the wide range spectrum, only the sharp peaks from the Te $(10\bar{1}0)$ lattice plane are observed. The micro-diffraction XRD was measured by Rigaku D/Max-RAPID II with Cu-Kα radiation (λ=1.5418 Å) at room temperature. The large curved imaging plate detector with a 210° aperture allows a two dimensional diffraction imaging over a broad $2\theta$ range, such that the detection of many lattice planes is possible without breaking the single crystal. As also shown in Supplementary Fig. S2a, the measured data matches well with the standard powder XRD pattern of Te. Supplementary Fig. S2b shows the high-resolution scanning transmission electron microscopy (STEM) (Thermo Fisher Themis G2 60-300) imaging and selective area electron diffraction (SAED) pattern of the exposed surface, which both further demonstrate the high quality of the single crystal at the atomic scale. The crystalline long axis, i.e., the growth direction, is the [0001] direction. The lattice parameters were determined to be a = b = 4.48 Å, c = 6.00 Å. The chemical composition of the crystals was examined by an energy dispersive spectrometer (EDS) (INCA detector, Oxford Instruments) attached to a field emission scanning electron microscope (FESEM) (Sirion 200, FEI) operated at 20 kV. No impurity elements were detected (Supplementary Fig. S2c).

**Magnetic and transport properties measurements.** The magnetic properties were analyzed using a Quantum Design SQUID-VSM system with the magnetic field up to 7 T and the temperature down to 2 K. For electrical transport measurements, Hall bar structures were fabricated via a standard e-beam lithography process, then 10 nm Pd and 50 nm Au were deposited as electrode materials using e-beam evaporation method. The measurements were performed in a Quantum Design PPMS system with the magnetic field up to 14 T.

**Angle-resolved photoemission spectroscopy (ARPES) measurements.** The ARPES measurements were performed at beamline 9U (Dreamline) of the Shanghai Synchrotron Radiation Facility (SSRF) with a Scienta Omicron DA30L analyzer. The angle resolution was 0.1°, and the combined instrumental energy resolution was better than 20 meV. The photon energy used in our experiments ranged from 25 to 90 eV, and the measurements were conducted at 7 to 20 K. The ($11\bar{2}0$) surface of the single crystals was cleaved, corresponding to the $k_y$-$k_z$ plane in the hexagonal Brillouin zone (BZ). The Fermi level of the Te samples was compared to a gold film reference which was evaporated onto the sample holder. All the measurements were conducted under a vacuum better than $7 \times 10^{-11}$ mbar.

**Electronic band structure calculations.** First principles calculations were carried out within the framework of density functional theory using the Vienna Ab initio Simulation Package (VASP)[34]. All the calculations were performed with a plane-wave cutoff of 450 eV on the 9×9×9 Γ-centered k-mesh, and the convergence criterion of energy was $10^{-6}$ eV. The generalized gradient approximation (GGA) with the Perdew-Burke-Ernzerhof (PBE) functional[35] was adopted to describe electron exchange and correlation. During structure optimization, all atoms were fully relaxed until the force on each atom was smaller than 0.01 eV/Å. In order to obtain a reliable band gap, the HSE06 hybrid functional[36] was used in the band structure calculation. The rotational symmetry $g = \{C_{2\hat{x}}|00\frac{2}{3}\}$ is a generator along the H-L line. Including the SOC, the rotational symmetry $g$ satisfies $g^2 = -1$, so its eigenvalues are $\pm i$. The first principles calculated eigenvalues for $g$ at two arbitrary momentums ($k_1$ and $k_2$) are shown in Fig. 1c. The eigenvalue is switched between two bands at $k_1$ and $k_2$, thus, there must exist a band crossing point with two-fold degeneracy between $k_1$ and $k_2$. Hence, the Weyl point of $W_1$ is protected by rotational symmetry. It is well known that the Kramers degeneracy is a two-fold degeneracy at time-reversal-invariant momentum (TRIM). Here, the Weyl point of $W_2$ is at L point, which is a TRIM. Hence, the Weyl point of $W_2$ is protected by time-reversal symmetry. The chirality of a Weyl point is calculated by integrating the Berry curvature on a surface enclosing the Weyl point.

**Acknowledgements**

This work was supported in part by the National Natural Science Foundation of China (Grants No. 11434009, U1832151, 11804326, 11774325 and 21603210), the National Key Research and Development Program of China (Grants No. 2017YFA0403600 and 2017YFA0204904), the Anhui Initiative in Quantum Information Technologies (Grant No. AHY170000), Hefei Science Center CAS (Grant No. 2018HSC-UE014), the Anhui Provincial Natural Science Foundation (Grant No. 1708085QA20), the Fundamental Research Funds for the Central Universities (Grant No. WK2030040087 and WK3510000007), and the Science, Technology, and Innovation Commission of Shenzhen Municipality (Grant No.KQTD2016022619565991).

**Author contributions**

C.Z. and L.L. conceived the project. N.Z. and L.L. grew the samples and performed the transport measurements with assistance from Z.L.; G.Z. and Z.W. performed the theoretical calculations; P.W. and Z.S. conducted the ARPES measurements; L.X. and J.H. performed the STEM experiments; L.L., N.Z., G.Z., Z.W., and C.Z analyzed the data and wrote the manuscript. All authors contributed to the scientific discussion and manuscript revisions.

## Competing interests

The authors declare no competing interests.

## Additional information

Correspondence and requests for materials should be addressed to C.Z. (cgzeng@ustc.edu.cn), Z.W. (zfwang15@ustc.edu.cn), or L.L. (lilin@ustc.edu.cn).

# Figures

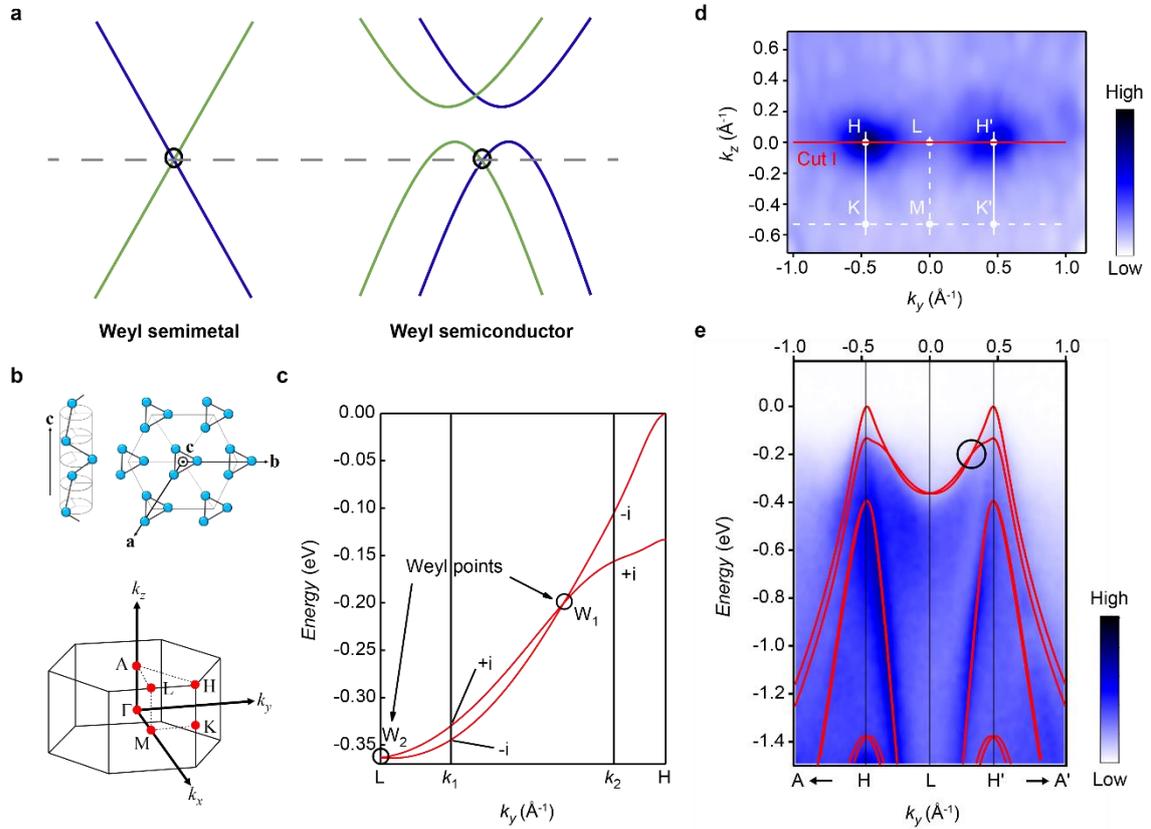

**Fig. 1 | Electronic band structure of tellurium. a,** Schematic conceptual illustrations of a Weyl semimetal and a Weyl semiconductor, respectively. Weyl points are indicated by the black circles. **b,** Crystal structure and Brillouin zone of trigonal Te. **c,** Calculated energy dispersion of electronic bands along the H-L line with SOC. The two Weyl points close to the top VBs are marked by the black circles, which have opposite chirality (+1 for $W_1$ and -1 for $W_2$). The calculated eigenvalues of the rotational operation at two arbitrary momentums ($k_1$ and $k_2$) are also marked (see Methods for more details). **d,** ARPES-intensity mapping of the Fermi surface for $k_x \sim \pi$ (MKHL plane). **e,** Comparison of the energy dispersion between ARPES experiments and theoretical calculations (red lines) along the H-L line.

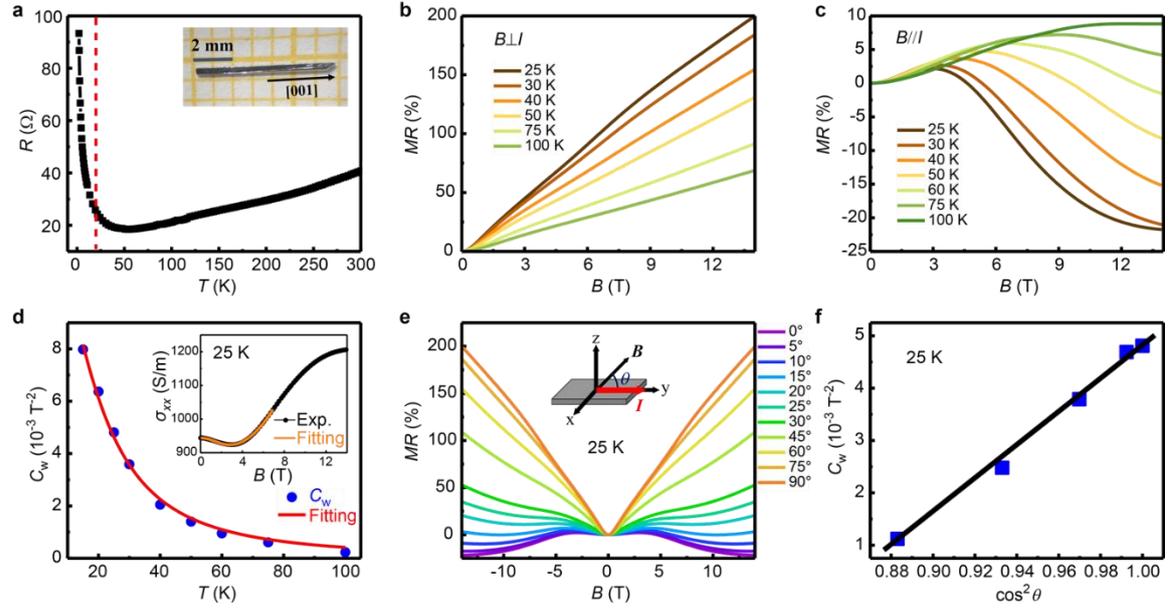

**Fig. 2 | Negative longitudinal magnetoresistance effect**. **a,** Temperature-dependent resistance of sample #3. The red dashed line indicates the rough boundary between the extrinsic region (T > 20 K) and the freeze-out region (T < 20 K). Inset: the optical image of the as-grown rod-like crystal, where the crystalline long axis lies along [0001]. **b,c,** MR measured at temperatures from 25 K to 100 K for $B$ perpendicular and parallel to $I$, respectively. **d,** Temperature dependence of extracted chiral coefficient $C_w$, which can be well fitted by using $C_w \propto v_F^3 \tau_v / (T^2 + \mu^2/\pi^2)$; inset: field dependence of the longitudinal conductivity at 25 K (black dots) and the corresponding fitting curve (orange line) using formula (1). **e,** MR measured at different angles ($\theta$) between $B$ and $I$. $\theta$ is better illustrated in the inset. **f,** Angular-dependent $C_w$ at 25 K, which is linearly dependent on $\cos^2\theta$.

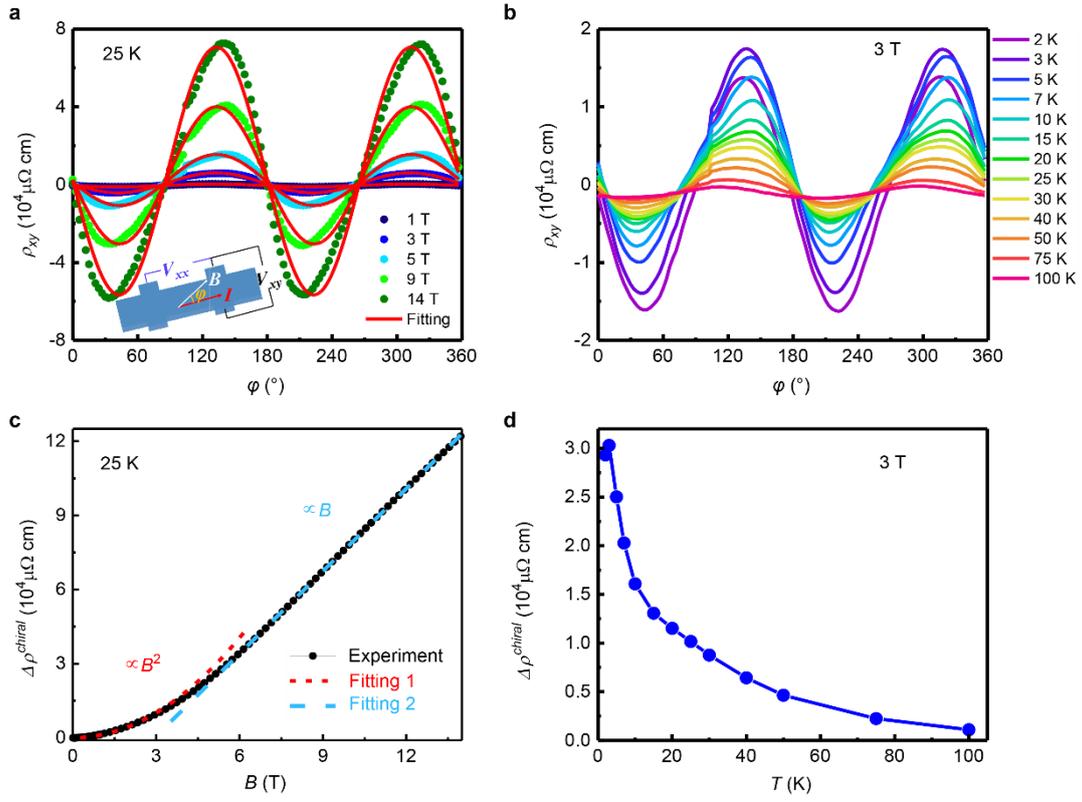

**Fig. 3 | Planar Hall effect. a,** Angular-dependent $\rho_{xy}$ measured under different $B$ at 25 K. The fitting curves using equation (4) are also plotted (red lines). The inset is the schematic of the PHE measurement geometry. **b,** Angular-dependent $\rho_{xy}$ measured at 3 T for different temperatures. **c,** Magnetic field dependence of the chiral-anomaly-induced anisotropic resistivity $\Delta\rho^{chiral}$ obtained at 25 K. In the low field regime, $\Delta\rho^{chiral}$ is proportional to $B^2$. **d,** Evolution of the extracted $\Delta\rho^{chiral}$ at 3 T as a function of temperature.

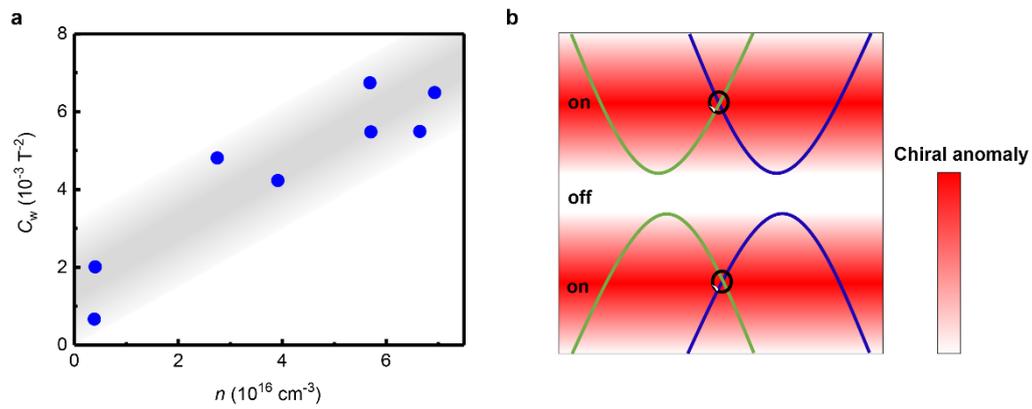

**Fig. 4 | Chemical potential dependent chiral anomaly in Weyl semiconductors. a,** The extracted chiral coefficient $C_w$ versus carrier density from the MR data of various Te samples. **b,** Schematic illustration of the evolution of the chiral anomaly with chemical potential.